\documentclass[draft,grl]{agu2001}
\usepackage{lineno}
\usepackage{graphicx}
\usepackage{amsmath}
\usepackage{amssymb}
\usepackage{xspace}
\usepackage[large]{subfigure}

\newcommand{\pdiff}[2][{}]{\ensuremath{\partial_{#2}\ifx\\#1\else^{#1}\fi}}

\newcommand{\uvec}[1]{\ensuremath{\mbox{\boldmath${\hat{#1}}$\unboldmath}}}
\newcommand{\unit}[2][{}]{\ensuremath{\ifx\\#1\else#1\,\fi\mathrm{#2}}}

\newcommand{\RS}{\mathrm{R_S}}

\newcommand{\RDISK}{\mathrm{R_{DISK}}}

\newcommand{\Cassini}{\textit{Cassini}\xspace}
\newcommand{\Voyager}{\textit{Voyager}\xspace}

\newcommand{\CAPS}{\renewcommand{\CAPS}{CAPS\xspace}\Cassini Plasma Spectrometer (CAPS)\xspace}
\newcommand{\MAG}{\renewcommand{\MAG}{MAG\xspace}\Cassini magnetometer (MAG)\xspace}
\newcommand{\MIMI}{\renewcommand{\MIMI}{MIMI\xspace}\Cassini Magnetospheric Imaging Instrument (MIMI)\xspace}
\newcommand{\INCA}{\renewcommand{\INCA}{INCA\xspace}ion-neutral camera (INCA)\xspace}
\newcommand{\PLS}{\renewcommand{\PLS}{PLS\xspace}\Voyager plasma science (PLS)\xspace}
\newcommand{\SLT}{\renewcommand{\SLT}{SLT\xspace}Saturn local time (SLT)\xspace}
\newcommand{\IBS}{\renewcommand{\IBS}{IBS\xspace}ion beam spectrometer (IBS)\xspace}
\newcommand{\IMS}{\renewcommand{\IMS}{IMS\xspace}ion mass spectrometer (IMS)\xspace}

\newcommand{\rev}[1]{\renewcommand{\rev}[1]{Rev~##1\xspace}Revolution~#1 (Rev~#1)\xspace}

\newcommand{\comment}[1]{}
\newcommand{\ZCS}{$Z_{\mathrm{CS}}$\xspace}
\newcommand{\ZSC}{$Z_{\mathrm{S/C}}$\xspace}
\newcommand{\ZCSMATH}{Z_{\mathrm{CS}}\xspace}
\newcommand{\ZSCMATH}{Z_{\mathrm{S/C}}\xspace}

\authorrunninghead{ACHILLEOS ET AL.}

\titlerunninghead{TITAN PLASMA ENVIRONMENT}

\authoraddr{N. Achilleos, 
Department of Physics and Astronomy, University College London,
Gower Street, London WC1E 6BT, UK. (nicholas.achilleos@ucl.ac.uk)}

\linenumbers*[1]

\begin{document}

%
%
%
%
%

%
%

\title{Sources of Pressure in Titan's Plasma Environment}
%

%
%




\authors{N. Achilleos, \altaffilmark{1}$^{^,}$\altaffilmark{3}$^{^,}$\altaffilmark{6} 
C. S. Arridge, \altaffilmark{2}$^{^,}$\altaffilmark{3}  
C. Bertucci, \altaffilmark{5}
P. Guio, \altaffilmark{1}$^{^,}$\altaffilmark{3} 
N. Romanelli, \altaffilmark{5}
N. Sergis \altaffilmark{4}  }

\altaffiltext{1} 
{Department of Physics and Astronomy, University College London, Gower Street, London, UK}

\altaffiltext{2} 
{Mullard Space Science Laboratory, Holmbury St. Mary, Dorking, Surrey, UK}

\altaffiltext{3} 
{Centre for Planetary Sciences at UCL / Birkbeck, University College London, Gower Street, London, UK}

\altaffiltext{4} 
{Office for Space Research and Technology, Academy of Athens, Athens, Greece}

\altaffiltext{5}
{Instituto de Astronom\'{i}a y F\'{i}sica del Espacio, University of Buenos Aires, Buenos Aires, Argentina}

\altaffiltext{6}
{Visiting professor at Japan Aerospace Exploration Agency
Institute of Space and Astronautical Science (ISAS), Sagamihara, JAPAN}

%
%
%

%
%


\begin{abstract}
In order to analyze varying plasma conditions upstream of Titan, we have
combined a physical model of Saturn's plasmadisk with a geometrical model
of the oscillating current sheet.
During modeled oscillation phases where Titan is furthest from the current sheet, the main
sources of plasma pressure in the near-Titan space are the magnetic pressure and, for disturbed conditions,
the hot plasma pressure. When Titan is at the center of the sheet, the main
source is the dynamic pressure associated with Saturn's cold, subcorotating plasma.
Total pressure at Titan (dynamic plus thermal plus magnetic) typically increases by a factor of five as the
current sheet center is approached. 
The predicted incident plasma flow direction deviates from the
orbital plane of Titan by $\unit[\lesssim10]{^{\circ}}$. These results suggest a correlation between the
location of magnetic pressure maxima and the oscillation phase of the plasmasheet.
\end{abstract}

%
%

%

\begin{article}

%
%

\section{Introduction}

%
%


%
%
Titan is usually
embedded within the rotating magnetosphere of Saturn - a configuration which leads to
the formation of a `flow-induced' magnetosphere, via the draping of the magnetic field  in the subcorotating
flow about the moon (Titan's orbital speed of $\unit[\sim6]{km\,s^{-1}}$ is small compared to the typical
speed of the rotating plasma, $\unit[\sim120]{km\,s^{-1}}$). 
Recently, \cite{bertucci2009} demonstrated, using \Cassini data, that
the direction and magnitude of the magnetic field upstream of Titan varies, depending mainly on whether
Titan is located above or below Saturn's magnetospheric current sheet.
Titan's distance from the current sheet is influenced by global magnetospheric 
oscillations at Saturn, which change the elevation of this structure with respect to the rotational equatorial plane.
The sheet geometry was modeled by  \cite{arridge2011} (hereafter A11). 

\begin{figure}[h]
\vspace{-4cm}
\noindent\includegraphics[width=25pc]{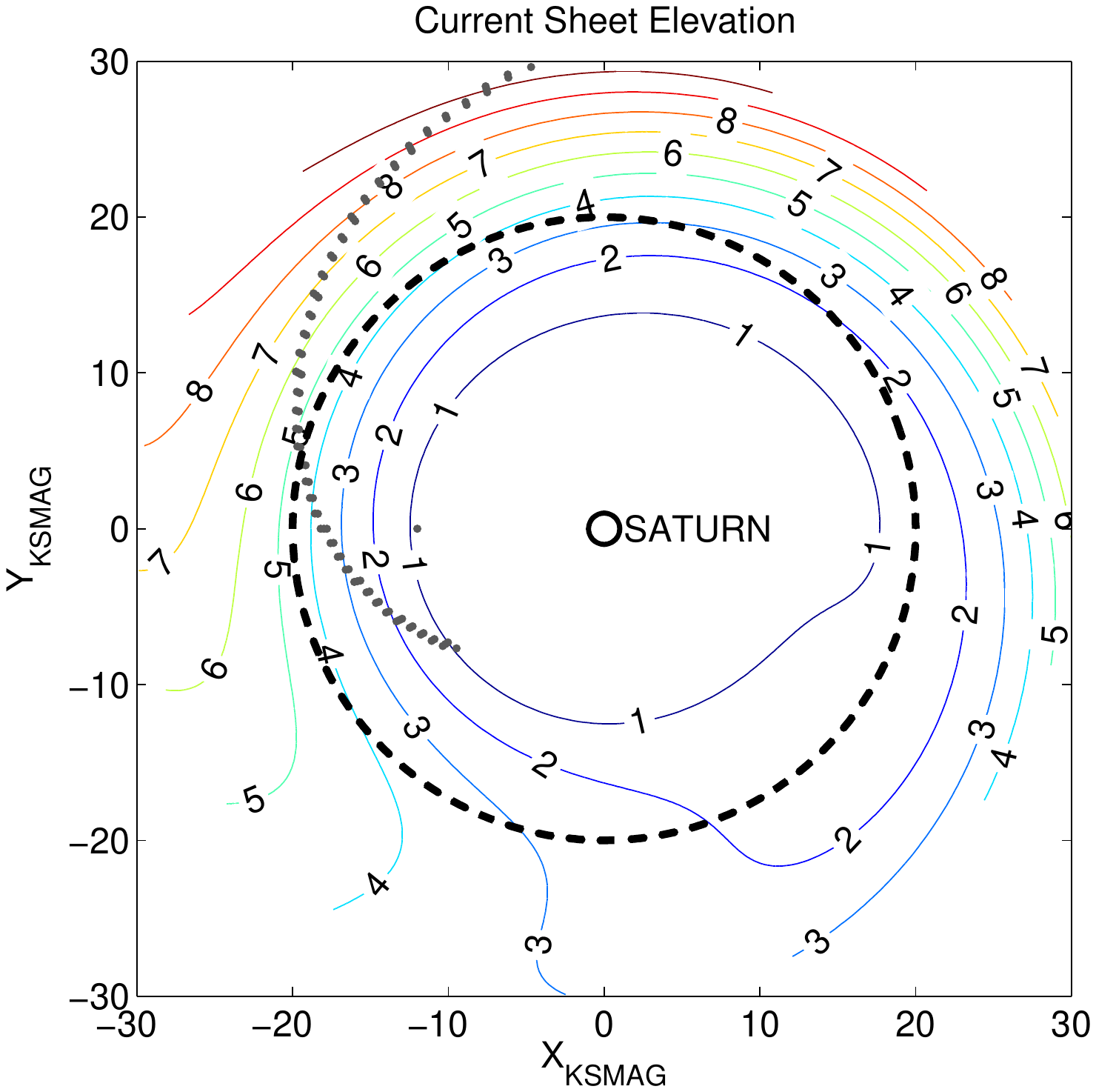}
\caption{
Plasma sheet geometry: 
Contours indicate the altitude \ZCS (in units of Saturn radii $R_S$)
of the A11 model current sheet above Saturn's rotational equator (see text).
The geometry shown is for southern
summer. The black, dashed circle is Titan's orbit and the gray squares represent a curve of constant `phase' in the sheet - this
curve passes through the point of maximum \ZCS at each radial distance. The $X_{\mathrm{KSMAG}}$ axis is the intersection of the
rotational equator and the noon meridian of Saturn local time (SLT). The whole pattern rotates with a variable period, 
following that of the SKR.
}
\label{fig:sheet}
\end{figure}

In
Figure \ref{fig:sheet}, we plot one example of current sheet elevation, \ZCS, from this A11 model. 
For constant radial distance (e.g.\ along Titan's orbit), \ZCS will vary
with azimuth - i.e.\ there is a `ripple' in the sheet surface. For southern summer, \ZCS is everywhere positive - hence, the
azimuthally averaged surface forms a `bowl-like' shape. We have combined the A11 model of sheet geometry with the Saturn plasmadisk
model of \cite{achilleos2010a} (hereafter Ach10) in order to predict the variable magnetic and plasma parameters during the T15 encounter of Titan by the \Cassini spacecraft (closest approach occurred on July 2, 2006 at 09:21 UTC, at altitude
\unit[\sim1900]{km}). This analysis enables us to predict the variations 
which arise from plasma sheet oscillations. In future, we aim to repeat the analysis for additional Titan encounters, and so provide 
a theoretical analog of observational classifications of the Titan
environment, such as those of \cite{rymertitan2009} and \cite{simon2010}. We have also chosen T15 for the present analysis 
because we have obtained relevant plasma moment data which we compare with our model results herein.

In \S\ref{sec:modinputs}, we describe a combined model which employs the A11 current sheet geometry with the 
Ach10 magnetic field / plasma model.
In section \S\ref{sec:modoutputs}, we implement this model and compare it to observations of the magnetic field, magnetic pressure and hot plasma pressure for several magnetospheric
oscillation periods centered on the T15 Titan flyby
(hot plasma pressure refers to H+ and O+ ions with energies $>\unit[3]{keV}$ \citep{sergis2009}).
We summarize and give conclusions in \S\ref{sec:summary}. 

\section{Plasmadisk Model Description}
\label{sec:modinputs}

For this study, we require a `two-component' model of Saturn's plasmadisk. The first component is the A11 geometrical model of the
current sheet, illustrated in Figure \ref{fig:sheet}. For cylindrical radial distance exceeding
\unit[\sim12]{\RS}, the altitude \ZCS of the current sheet (with respect to Saturn's rotational equator) is given by A11:

\begin{align}
Z_{\mathrm{CS}}  =  [(\rho - R_H\,\tanh(\rho/R_H)] \tan(-\theta_{\odot}) 
                             +  \tan(\theta_T)\,(\rho-\rho_o)\,\cos(\lambda), \hspace{1cm} \rho > 12{\mathrm{R_S}}
\label{eq:zcsdefn}
\end{align}

\noindent where the first term represents the axisymmetric bowl and the second term the
spatial oscillation, or ripple. Symbols have the following meanings: $\rho$ is cylindrical
radial distance with respect to the planet's rotational / magnetic axis, $R_H$ is the hinging
distance which controls the curvature of the bowl, $\theta_{\odot}$ is the
subsolar latitude at Saturn, $\theta_T$ is an effective angle of tilt for the current sheet, $\rho_o=12\RS$ is
a scaling distance which controls the amplitude of the ripple. $\lambda$ represents
the following phase angle for describing plasma sheet oscillation, dependent on both position and time:

\begin{align}
\lambda = \lambda_{SLS3} - \lambda_o - \Omega_{SKR}\,(\rho-\rho_o)/V_{WAVE},
\end{align}

\noindent where SLS3 denotes the longitude of \cite{kurth2008}, based on fitting a low-order polynomial
to the non-stationary period of the Saturn Kilometric Radiation (SKR). Since SLS3 was developed, distinct SKR signals have
been identified in Saturn's northern and southern hemisphere (e.g.\ \cite{lamy2011}) - the SLS3 phase lies consistently 
within $\unit[\sim30]{^{\circ}}$ of
that of the southern SKR source \citep{andrews2010}. $\lambda_o$ denotes a `prime meridian' parameter, fitted by A11 to
different passes of \Cassini data. $\Omega_{SKR}$ is a variable angular velocity corresponding to the SLS3 period. $V_{WAVE}$
is a `wave speed' parameter which introduces a systematic delay of the oscillation phase with radial distance (see Figure \ref{fig:sheet}).
The T15 Titan encounter occurred during \Cassini's Revolution 25. We thus adopt the same sheet parameters as used by 
A11 for their Rev 25 model fit, namely 
$\lambda_o = 100^{\circ}$, $V_{WAVE} = 5\, \unit[\RS]{hr^{-1}}$, $R_H = 16 \,\RS$, $\theta_T = 12^{\circ}$, $\rho_o = 12 \, \RS$.



The second component of our plasmadisk model specifies magnetic field and plasma distributions for an axisymmetric
magnetosphere in which magnetic force, centrifugal force and plasma pressure forces are in equilibrium
(Ach10). This model also assumes north-south symmetry, with a current sheet lying in the
rotational equator. Any plasma parameter is a function of two  coordinates,  labeled
$\rho_{\mu}$ and $Z_{\mu}$, the respective cylindrical radial distance and altitude (with respect to the rotational equator) in the `Ach10 model
space'. In order to combine the Ach10 model with the A11 sheet geometry, we calculate `equivalent Ach10 model
coordinates' corresponding to the spacecraft's actual location:

\begin{align}
\rho_{\mu} = \rho_{S/C}  \notag  \\
Z_{\mu} = (\ZSCMATH-\ZCSMATH)\,\uvec{z}\cdot\uvec{n},
\label{eq:coordtrans}
\end{align}

\noindent where $\rho_{S/C}$ is the spacecraft's actual cylindrical radial distance from the planet's rotation / dipole axis,
\ZSC and \ZCS are the respective altitudes of the spacecraft and the A11 current sheet with respect to the rotational equator,
$\uvec{z}$ is a unit vector pointing in the northern direction of the planet's axis, and $\uvec{n}$ is the unit vector normal to the A11 
current sheet at the distance $\rho = \rho_{S/C}$. These expressions assume that the local structure of the
plasmadisk (at \Cassini) may be approximated by a version of the Ach10 model, whose plane of symmetry has been rotated to match the local 
tangent plane of the A11 sheet.

\section{Comparison of Plasmadisk Models and T15 Observations}
\label{sec:modoutputs}
\begin{figure}[h]
\noindent\includegraphics[width=30pc]{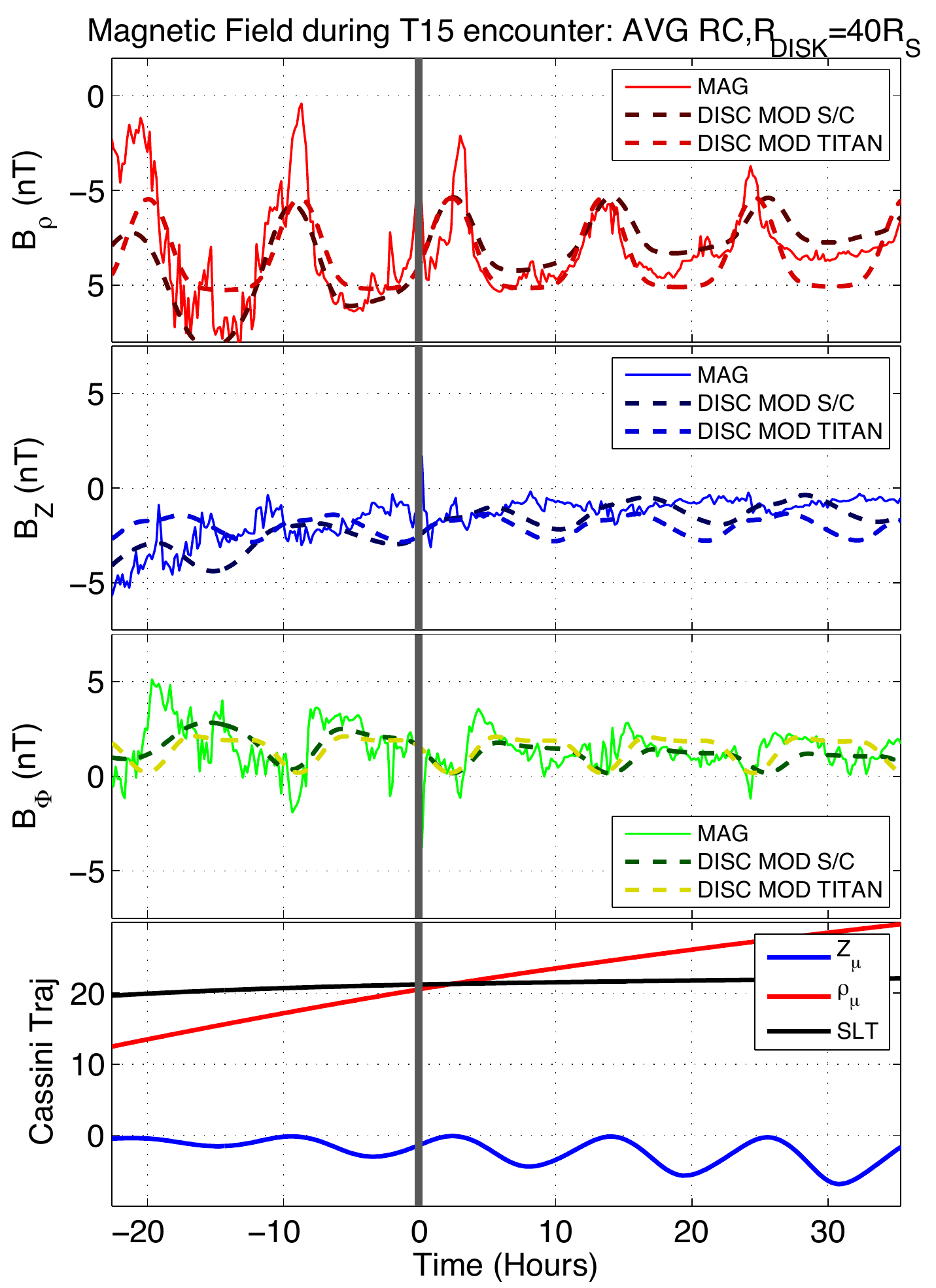}
\caption{
Top three panels: Cylindrical components of the magnetic field observed by
\Cassini, and predicted by the model, during several magnetospheric oscillations before and after the T15
wake crossing (vertical gray line). The zero of time indicates closest approach to Titan. Model fields for
both \Cassini and Titan-based observers are shown. Bottom panel: Equivalent Ach10 model coordinates
along the spacecraft trajectory. $Z_{\mu}$ indicates perpendicular distance from the spacecraft to the A11
current sheet.
}
\label{fig:T15BField}
\end{figure}

In Figure \ref{fig:T15BField}, we show the observed and modeled components of the magnetic field
in cylindrical coordinates. The two Ach10 model parameters explored, in order to fit the data, are the effective
magnetodisk radius $\RDISK$ and a proxy for the ring current activity which makes use of the global hot plasma pressure,
based on multi-orbit statistics of the pressure moments from the \MIMI instrument
(see \cite{achilleos2010b,sergis2007}). The fit shown is for $\RDISK=40\,\RS$ and average
ring current (equivalent to hot plasma pressure $P_H = \unit[2\times10^{-3}]{nPa}$ at Titan's orbit). The r.m.s.\
difference between the data and model (summed over radial and azimuthal components, for the time interval plotted) changes by
$\sim30\%$ for corresponding changes $\Delta\RDISK=5\,\RS$ and $\Delta P_H =  \unit[2.5\times10^{-4}]{nPa}$.
We show several
magnetospheric oscillations. The fits to the amplitude and phase
of the $B_{\rho}$ (radial) and $B_{\phi}$ (azimuthal) fields are reasonable, although:
(i) Earlier $B_\rho$ data show a change in sign, indicative of passage north of the current sheet plane, which
is not reproduced with the model; (ii) The $B_{\phi}$ fluctuations show a much steeper
`rising' part compared to the model, suggesting that the plasmasheet ripple exhibits structure more complex than
a sinusoidal form (equation~\ref{eq:zcsdefn}). The model $B_Z$ is almost in
antiphase with the observation, also suggesting additional plasmasheet structure beyond our 
`wavy disk' model  (e.g. a rotating azimuthal anomaly in hot pressure has been proposed 
by \cite{brandt2010}).

\begin{figure}[h]
\noindent\includegraphics[width=19pc]{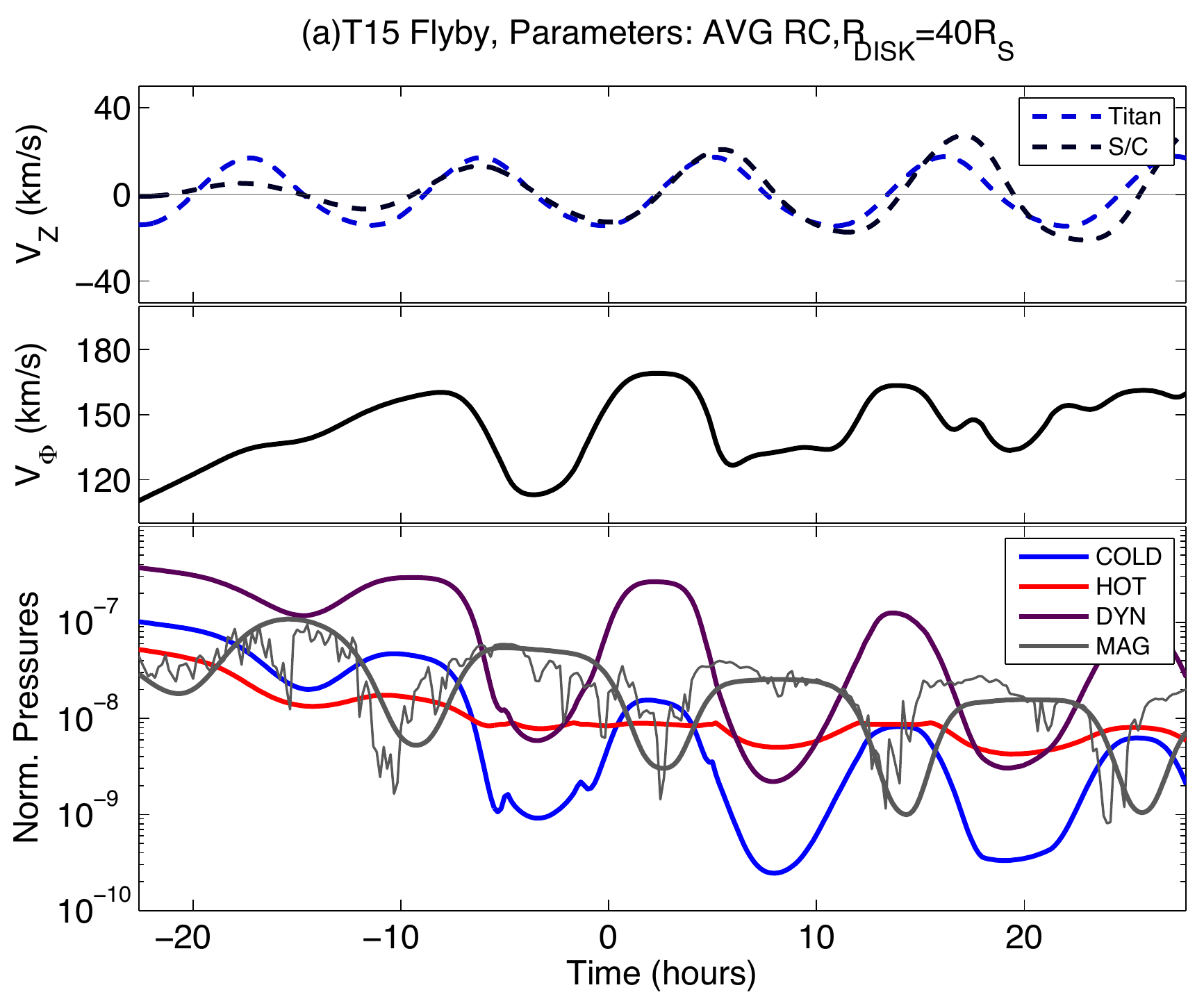} 
\noindent\includegraphics[width=19pc]{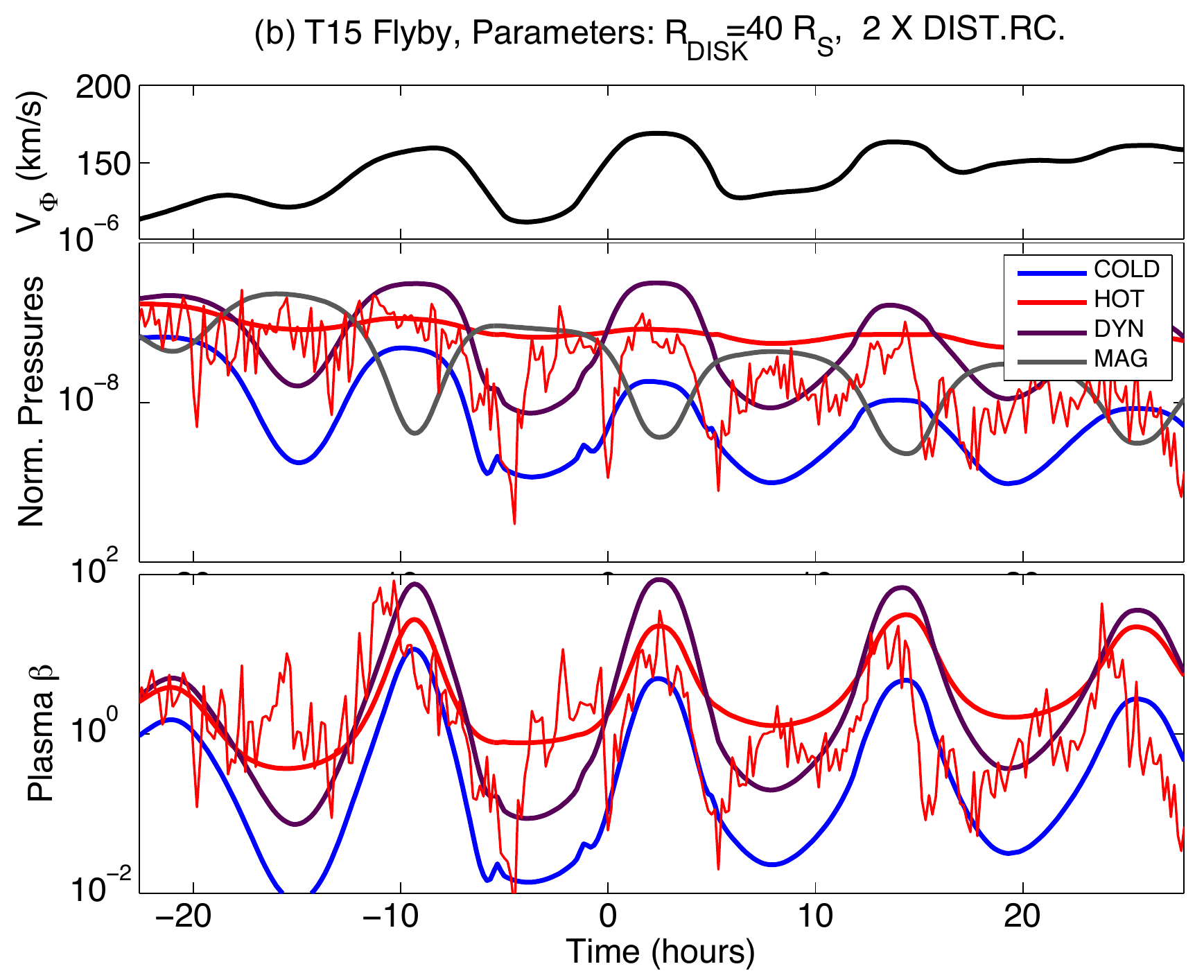}
\caption{
(a) Model predictions for a disk of effective radius $\RDISK=40\RS$ and average
ring current level (see text).
Top panel: Predicted vertical velocity components for the plasmasheet for the same
time interval as Figure \ref{fig:T15BField}. Middle panel: Model azimuthal velocities for
the cold plasma on planetary flux tubes conjugate to the spacecraft. Bottom panel:
Predicted pressure contributions, color-coded according to physical origin.
Pressure is normalized to the value of magnetic pressure corresponding to
the equatorial surface field strength at Saturn, $B_o = \unit[21000]{nT}$.
(b) Model predictions for a disk of effective radius $\RDISK=40\RS$ and twice the
hot pressure for the `disturbed' ring current level.
Top panel: Azimuthal velocities for
the cold plasma on planetary flux tubes conjugate to the spacecraft. Middle panel:
Pressure contributions, color-coded, as for Figure \ref{fig:T15Plasma}a, according to physical origin (see text).
Bottom panel: Plasma beta parameters corresponding to the model pressures
and to the observed hot plasma / magnetic pressure.
}
\label{fig:T15Plasma}
\end{figure}

In Figure \ref{fig:T15Plasma}a, we show model plasma parameters corresponding to the field model of
Figure \ref{fig:T15BField}. The vertical velocity $V_Z$ of the plasmasheet is similar for Titan and \Cassini 
reference frames near closest approach, with values up to $\unit[\sim30]{km\,s{-1}}$.
Similar vertical velocities were measured by the \Cassini plasma spectrometer (CAPS) during the T15
flyby \citep{sillanpaa2011}.
The azimuthal velocity of the cold plasma, lying on field lines conjugate with the spacecraft, is shown in the middle panel.
The largest northward excursions of  the plasmasheet (zero-crossing points which occur 
after the positive maxima in $V_Z$) are accompanied by decreases in $V_{\phi}$ as the spacecraft moves away
from the current sheet, and connects to
flux tubes extending to larger radial distances, which rotate more slowly.
Note that $V_{\phi}$ for the interval shown, combined with the 
$V_Z$ for the Titan frame, indicate that the upstream plasma 
flow direction is tilted with respect to the rotational equator by angles 
$\lesssim12^{\circ}$. The location of maximum magnetic pressure along
draped flux tubes would also be expected to deviate from the rotational equator, for appropriate
oscillation phases.

The observed magnetic field is 
dominantly radial outside the current sheet. The maximum value of
$|B_{\rho}/B_{Z}|$ for the interval shown is $\sim20$, which also equals the maximum ratio
$|E_Z / E_{\rho}|$ for the convective electric field (see \cite{arridge2011b} for more details).

In the bottom panel of Figure \ref{fig:T15Plasma}a, we show the contributions to plasma pressure from
various sources. The maximum pressure
during current sheet encounters is provided by the dynamic pressure of the cold,
subcorotating plasma (violet curve). In the exterior regions or `lobes' of the sheet, magnetic pressure shows local
maxima and is the dominant pressure source for this average-ring-current model.
The amplitudes and
phasing of the observed
fluctuations in magnetic pressure (thin gray curve) are in reasonable agreement with the model - although the narrower observed minima
suggest a thinner sheet. The hot plasma pressure (red curve)
shows relatively weak fluctuations compared to the other curves, since we assume that the
hot population has uniform pressure all the way along the field lines. The blue curve indicates 
thermal pressure of the cold plasma. The total 
effective pressure predicted by the model (i.e.\ dynamic
plus thermal plus magnetic) typically increases by a factor of approximately five as
the current sheet center is approached. This change is mainly due to the variability in dynamic pressure
between the sheet center and lobes (the relative change in pressure becomes 
$\sim10\%$ if dynamic pressure is excluded).

In Figure \ref{fig:T15Plasma}b, we compare observed and modeled 
hot plasma pressure. The 
global profile of hot plasma pressure was increased to twice the value for the 
`disturbed ring current'  of A10. Comparing the red curves (middle panel), 
the observed hot plasma pressure outside the current sheet 
(local maximum in magnetic pressure) is in agreement with the model. The data show additional variability in hot
pressure, due to plasma injections and ion beams, which are not explicitly modeled.
This disturbed-ring-current model, in comparison to the average-ring-current case
(Figure \ref{fig:T15Plasma}b),  shows more comparable values of magnetic
and hot pressure in the lobes of the sheet.

The bottom panel of Figure \ref{fig:T15Plasma}b shows model plasma beta parameters, and the observed hot plasma beta from the
\Cassini data. The model hot plasma beta ($\beta_h$) varies between $\sim1-100$, while the observed $\beta_h$
reaches values as low as $\sim 0.01$.  A `pseudo' plasma beta may be defined for the dynamic pressure (Ach10)
according to $\beta_d = P_{\mathrm{dyn}} / P_{\mathrm{mag}}$, where subscripts indicate dynamic and magnetic pressures.
$\beta_d$ shows local maximum values comparable to those for $\beta_h$. The thermal cold plasma beta, 
$\beta_c$, shows the lowest values, down to $\sim10\%$  of
$\beta_h$. The ratio $\beta_c/\beta_d$ has similar minimum values, since the bulk kinetic
energy of the cold plasma ions far exceeds their thermal energy.

\section{Conclusions}
\label{sec:summary}
We have calculated a plasmadisk model for conditions at the orbit of Titan during the T15 encounter by \Cassini. 
Our model reproduces some of the
large-scale variability in the observed magnetic field, although more complex structure for the ripple in the
current sheet is required for better agreement. The model outputs are in reasonable agreement with the 
\Cassini observations of magnetic pressure and hot plasma pressure.

For magnetospheric oscillation phases where Titan is furthest from the current sheet, the field is strongly radial and the dominant source of
pressure is the dynamic pressure of the subcorotating, cold plasma. For phases where Titan is near the center of the sheet, the dominant pressure
sources are the magnetic and hot plasma pressures (the latter being prevalent for the disturbed ring current of T15).
Magnetospheric oscillations also control changes in vertical and azimuthal velocities of the cold plasma for a Titan-based observer. 
In our model, the incident direction of cold plasma flow may be displaced from Titan's orbital plane by angles of the order $\sim10{^{\circ}}$. This result is in good agreement with observations of the plasma flow velocity
by \cite{sillanpaa2011}.

Finally, the plasmadisk oscillations lead to a wide range of plasma beta regimes in which Titan may be immersed.
The hot plasma beta may be as high as $\sim100$ for phases when Titan is at the center of the disk.
The cold plasma beta is lower by factors of $\sim5-10$. A `pseudo' plasma beta 
associated with the cold plasma bulk rotation (dynamic pressure) exceeds even the hot plasma beta near the disk center.

This variability in plasma conditions presents a complex requirement for upstream boundary conditions, such as those used in more
sophisticated MHD models of the plasma flow. Our model is also useful for predicting plasma moments, when 
observations of these are scarce, or when only magnetic observations are available.



%
%

%
%

\begin{acknowledgments}
We acknowledge the continued collaboration of the \Cassini magnetometer (MAG) and
plasma (CAPS, MIMI) instrument teams. NA was supported by both a JAXA Visiting Professorship
and UK STFC Consolidated Grant ST/J001511/1 (UCL Astrophysics). CB acknowledges the financial support of
the Europlanet Visiting Researcher programme.
\end{acknowledgments}

%
%
%
%
%
%
%
%


\bibliographystyle{agu}
\bibliography{References}

\end{article}

\end{document}